\documentclass[12pt,preprint]{aastex}
\newcommand{\nraoblurb}{The National Radio Astronomy Observatory is
a facility of the National Science Foundation operated under cooperative
agreement by Associated Universities, Inc.}
\newcommand{\m}{$\,{\rm m}$}
\newcommand{\kpc}{$\,{\rm kpc}$}

\newcommand{\mhz}{$\,{\rm MHz}$}
\newcommand{\ghz}{$\,{\rm GHz}$}
\newcommand{\percc}{$\,{\rm cm^{-3}}$}

\newcommand{\kms}{${\,{\rm km\, sec^{-1}}}$}

\newcommand{\arcmper}{\rlap.{^{\prime}}}
\newcommand{\arcsper}{\rlap.{^{\prime\prime}}}

\newcommand{\mjy}{\,mJy}

\newcommand{\microjy}{\,$\mu$Jy}
\newcommand{\microjyb}{$\rm \,\mu Jy\,beam^{-1}$}
\newcommand{\h}[1]{$^{#1}{\rm H}$}
\newcommand{\car}[1]{$^{#1}{\rm C}$}
\newcommand{\li}[1]{$^{#1}{\rm Li}$}
\newcommand{\he}[1]{$^{#1}{\rm He}$}
\newcommand{\hii}{H~{$\scriptstyle {\rm II}$}}
\newcommand{\hep}[1]{$^{#1}{\rm He}^{+}$}
\newcommand{\hepr}[1]{$^{#1}{\rm He}^{+}/{\rm H}^{+}$}

\newcommand{\heppr}[1]{$^{#1}{\rm He}^{++}/{\rm H}^{+}$}
\newcommand{\her}[1]{$^{#1}{\rm He}/{\rm H}$}
\newcommand{\lir}[1]{$^{#1}{\rm Li}/{\rm H}$}
\newcommand{\hal}{${\rm H}\,91\alpha$}
\newcommand{\heal}{${\rm He}\,91\alpha$}
\newcommand{\heps}{${\rm H}\,154\epsilon$}
\newcommand{\heta}{${\rm H}\,171\eta$}
\newcommand{\halpha}{${\rm H}\,\alpha$}

\newcommand{\expo}[1]{${10^{#1}}$}
\newcommand{\nexpo}[2]{${#1 \times 10^{#2}}$}
\newcommand{\ngc}[1]{NGC\thinspace #1}
\newcommand{\jon}[1]{J\thinspace #1}
\newcommand{\etab}{$\eta_{\rm \,b}$}

\begin{document}
\title{The Detection of $^{\bf 3}$He$^{\bf +}$ in a Planetary Nebula
  Using the VLA}

\shortauthors{Balser et al. 2005}

\author{Dana S. Balser\altaffilmark{1}, W. M. Goss\altaffilmark{2},
T. M. Bania\altaffilmark{3} \& Robert T. Rood\altaffilmark{4}}

\altaffiltext{1}{National Radio Astronomy Observatory,
P.O. Box 2, Green Bank WV 24944, USA.}
\altaffiltext{2}{National Radio Astronomy Observatory,
Socorro, NM 87801, USA.}
\altaffiltext{3}{Institute for Astrophysical Research,
725 Commonwealth Avenue, Boston University, Boston MA 02215, USA.}
\altaffiltext{4}{Astronomy Department, University of Virginia,
P.O.Box 3818, Charlottesville VA 22903-0818, USA.}

\begin{abstract}

We used the VLA to search for \hep{3}\ emission from two Galactic
planetary nebulae (PNe): \ngc{6572} and \jon{320}.  Standard stellar
models predict that the \her3\ abundance ratios for PNe should be 1--2
orders of magnitude higher than the primordial value (\her3\ $\sim$
\expo{-5} by number) determined from Galactic \hii \ region abundances
and confirmed by WMAP cosmic microwave background results.  Chemical
evolution models suggest that fewer than 5\% of all PNe enrich the
interstellar medium (ISM) with \he3\ at the level of standard stellar
models. Our target PNe are therefore anomalous in that they were selected
from a sample deliberately biased to contain objects with properties
that maximized the likelihood of a \he3\ detection by the VLA. We have
detected the 8.665\ghz\ hyperfine \hep3\ transition in \jon{320} at
the $4\,\sigma$ level. The \her3\ abundance ratio is \nexpo{1.9}{-3}
with roughly a factor of two uncertainty. For \ngc{6572} we find an
upper limit of \her3\ $\la$ \expo{-3} .  This detection of \he3\ in
\jon{320} makes it the second PN known to have an anomalously high
\he3\ abundance confirming that at least some low-mass stars produce
significant amounts of \he3\ that survives to the PN stage and enriches
the ISM.
\end{abstract}


\section{Introduction}

The light elements, \h2 (D), \he3, \he4\ and \li7, are expected to
be produced during the era of primordial nucleosynthesis and then
further processed by stars over many generations \citep{reeves74,
boesgaard85}.  Observations of the light elements can in
principle be used to constrain models of stellar and Galactic
evolution and cosmology.  For each light element the quantity of
interest is the abundance ratio relative to hydrogen (e.g., \her3).
In some cases converting the observables into an abundance ratio can
be as difficult as making the observations.

\begin{figure}
\includegraphics[angle=-90, scale=0.7]{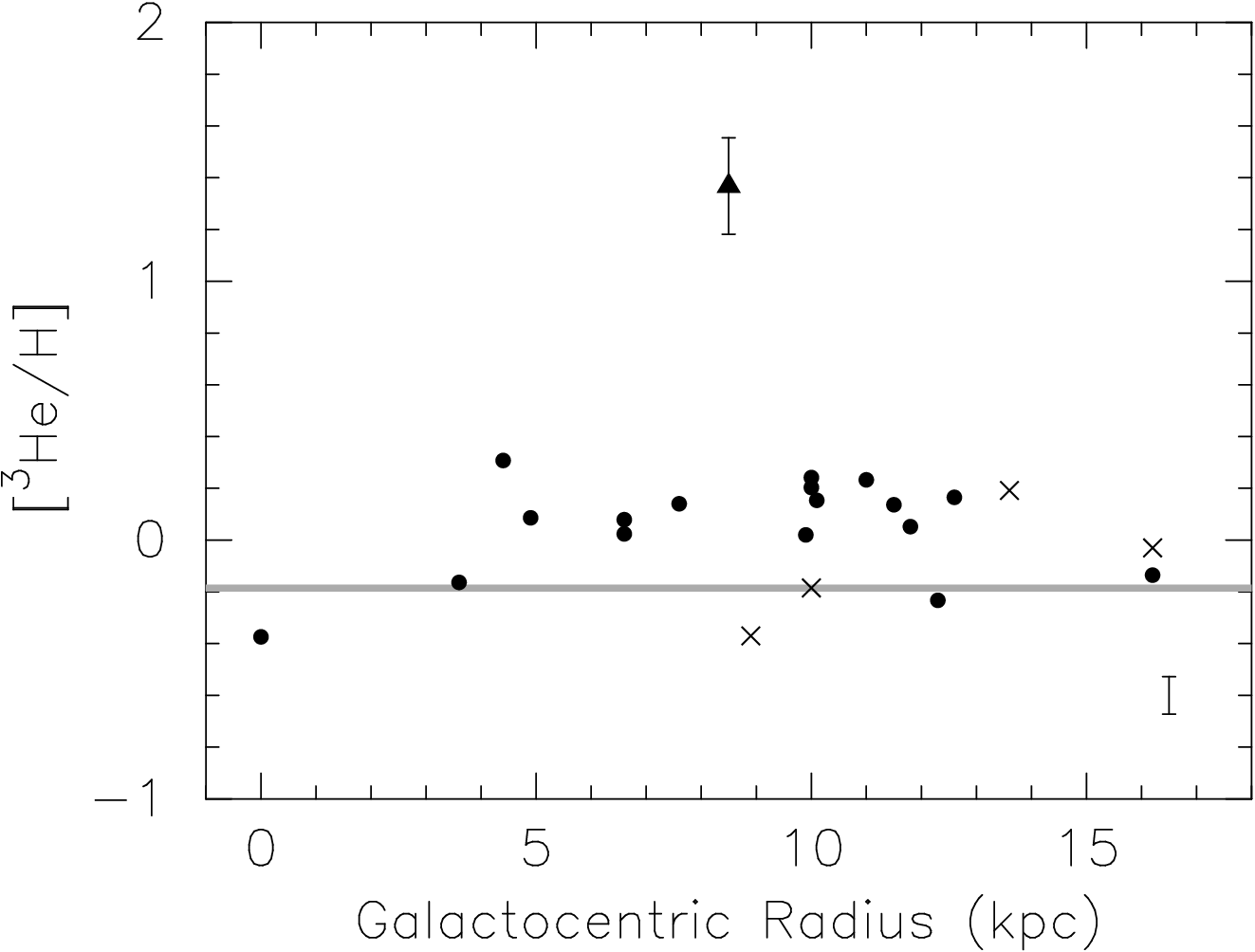}
\figcaption{
\her3\ abundance as a function of Galactocentric radius where
[\her3] = log(\her3) -- log($1.5 \times 10^{-5}$).  The solid points
and crosses are Galactic \hii\ region \her3\ abundance ratios taken
from \citet{bania02}.  The crosses correspond to \hii\ regions where
no ionization correction could be made and thus we may have
underestimated the \her3\ abundances in these objects.  The typical
error is shown in the lower right hand corner.  The triangle plots 
[\her3] for the planetary nebula \ngc{3242} \citep{balser97}.  
The thick solid line indicates the primordial \her3\ abundance predicted
by Big Bang nucleosynthesis \citep{burles01} assuming the WMAP 
baryon-to-photon ratio \citep{spergel03}.
\label{fig:he3}}
\end{figure}

For many years the light elements have been used to constrain standard
Big Bang nucleosynthesis (SBBN) models \citep{copi95}.  SBBN contains
only one free parameter, the baryon-to-photon ratio, \etab. Observations 
of optical recombination lines of \he4\ in metal poor galaxies were first
used to probe SBBN \citep{peimbert74}.  Observations of the resonance
line of \li7\ in metal poor Halo stars revealed a constant \lir7\
abundance ratio, called the ``Spite Plateau'', that was interpreted as
the primordial abundance and used to determine values of \etab\
\citep{spite82}.  The advent of large aperture, optical telescopes provided the
sensitivity to detect the Lyman-series lines of D in metal
poor QSO absorption systems \citep{tytler96}.  More recently the
\hep3\ hyperfine transition was used to estimate \etab\ by measuring
\her3\ in \hii\ regions throughout the Galaxy.  A ``\he3\ Plateau''
was discovered and from it a primordial abundance inferred
\citep[and shown in Figure~\ref{fig:he3}]{bania02}.  Since
the primordial D/H abundance ratio is more sensitive to \etab, and the
processing of D in stars is thought to be understood, it became the
baryometer of choice by many investigators \citep[for example]{schramm98}. 
Nonetheless, {\it all} the light element abundances should agree with SBBN 
within the uncertainties.  

Observations of the cosmic microwave background (CMB) provide an
independent method of determining \etab.  CMB anisotropy measurements
made with the Wilkinson Microwave Anisotropy Probe (WMAP) were used by
Spergel et al. (2003) to derive the primordial baryon-to-photon ratio.
Adopting the CMB results for \etab\ gives the primordial light element
abundances \citep{romano03, cyburt03, coc04, it04, steigman05}.  The D
and \he3\ abundances are in good agreement with those derived from the
WMAP observations.  Depending on the adopted uncertainties the
measured \he4\ abundance is either in agreement with CMB data or
slightly lower than expected \citep{peimbert03, olive04}. The \li7\
abundance is significantly lower given the low dispersion of these
measurements.  A possible resolution of this result is that the \li7\
abundances are depleted in metal poor Halo stars \citep[and references
within]{charbonnel05}.

\placefigure{fig:he3}

Figure~\ref{fig:he3} shows the \her3\ abundance ratio by number as a
function of Galactocentric radius \citep{bania02}.  The solid circles
(and crosses) are \he3\ abundances for \hii\ regions located
throughout the Galaxy.  Since \hii\ regions are very young, they
measure the \her3\ abundance ratio at the present time.  Not plotted
are the \her3\ abundances for protosolar material and the local ISM
derived from the {\sl Galileo Probe} \citep{mahaffy98} and {\sl
Ulysses} \citep{gloeckler98}, respectively, that are in accord with
the \hii\ region \he3\ abundances.  The solid triangle is the \he3\
abundance for the planetary nebula \ngc{3242}.  Early stellar and
Galactic evolution models predicted that solar type stars would
produce and eventually expel into the ISM prodigious amounts of \he3,
creating an easily measurable abundance gradient with Galactocentric
radius or metallicity \citep{rood76}.  The high \he3\ abundance in
\ngc{3242} was consistent with these early models but inconsistent with
the \hii\ region \he3\ abundances, stimulating a new generation of
models.

In standard stellar models \he3\ is produced on the main sequence
outside the main nuclear burning regions where p-p burning terminates
in \he3. Longer lived low mass stars produce more \he3. This \he3\ is
mixed into the convective envelope at the first dredge up on the lower
red giant branch. Once in the envelope the \he3\ survives, so any
stellar winds and the final PN should be substantially enriched in \he3.

Standard stellar models include only mixing produced by thermal
convection. Observations have long indicated that {\em extra mixing}
processes must be taking place, e.g., overshooting beyond convective
radiative boundaries or rotationally induced turbulence
\citep{zahn92}. Charbonnel et al. (1998) argued that this rotationally
induced extra-mixing occurs in low mass stars above the luminosity
function bump produced when the hydrogen burning shell crosses the
chemical discontinuity left by the retreating convective envelope
early in the red giant branch phase. This hypothesis was supported by
observed increases in the ${\rm {}^{13}C}$ abundance above the
luminosity function bump. The same mixing that would increase the
${\rm {}^{13}C}$ abundance would bring the \he3\ down to values
comparable to the primordial value. Further extra-mixing could take
place on the second ascent asymptotic giant branch. In stars which
have undergone extra mixing the material expelled into the ISM during
the PN stage is therefore not enriched in \he3\ as originally thought.
\citet{charbonnel98b} predict that 96\% of all low-mass stars will
undergo this mixing process.  The expected \he3\ yields coupled with
Galactic evolution models \citep{tosi00, chiappini02, galli05}
are consistent with the ``\he3\ Plateau'' shown in Figure~\ref{fig:he3} and 
described by \citet{bania02}.

The horizontal grey band in Figure~\ref{fig:he3} is the \her3\
primordial abundance ratio predicted by SBBN calculations
\citep{burles01} using the WMAP \etab\ value.  The CMB
abundance lies at the lower envelope of the \her3\ abundances. 
This and the absence of an observable negative gradient in the \he3\ 
abundances with galactocentric radius led \citet{bania02} to argue
that the stellar contribution to \he3\ is
positive definite but very small. The scatter in the \hii\ region
abundances could be real (e.g., azimuthal abundance fluctuations).
But this scatter could also be measurement error (e.g., the ionization
correction could be underestimated for some \hii\ regions). The
scatter presumedly obscures a yet to be observed gradient. Given this,
a limit on BBNS production can be determined from the best determined
abundance in the outer Galaxy. The rightmost solid point in
Figure~\ref{fig:he3}, the \hii\ region S209, fits this requirement,
and \citet{bania02} took the S209 abundance as their best estimate of
the primordial abundance. This value is almost identical to the WMAP
value.

There are two observational paths to improving our understanding of
the evolution of \he3. The first is to reduce the scatter in the \hii\
region abundances by observing a sample chosen because of their simple
physical structure and optimum location in the Galaxy. The second is
to observe the stellar sources of \he3 and directly measure the \her3\ 
abundance in PNe.

Previous observations of six PNe were made using the Max Planck
Institut f\"ur Radioastronomie 100\m\ telescope \citep{balser97}.  The
PN sample was chosen based on criteria that would maximize detection,
including diagnostics that indicated less mixing.  \hep3\ was only
detected in \ngc{3242}, shown as the solid triangle in
Figure~\ref{fig:he3}.  It therefore appears that in at least one
object significant amounts of \he3\ in a low-mass star survives until
the PN phase, and that PN and its earlier stellar winds have enriched \he3\
in the ISM.  We now seek to understand why some
stars produce \he3\ and others do not. 

Observations of \hep3\ in PNe are challenging because of the weak
intensity of the \hep3\ emission line coupled with their small angular
size.  We are increasing the size of our PN sample using the Very Large 
Array (VLA), the 100\m\ Green Bank Telescope (GBT), and the 305\m\ Arecibo 
telescope.
Our target sample is purposely biased to maximize detections. 
Our target selection criteria have evolved with time. Currently they
are:

\begin{itemize}

\item Any mixing in the progenitor star beyond that included in
standard stellar models is likely to destroy \he3.  Indications of
mixing include: high \car{13}/\car{12}, N/O, and \he4/H\ abundances. 
To be included as one of our targets a PN must have a ${\rm ^{4}He/H}$ 
abundance ratio $< 0.125$ by number and ${\rm [N/O]} < -0.3$ with a 
value $< -0.6$ considered even better. Unfortunately the \car{13}/\car{12} 
abundance ratio is known for only a small sample of PNe
\citep{clegg97, palla00, palla02, balser02}.

\item The \he3\ abundance in the progenitor grows as main sequence lifetime 
increases, hence PNe from older stellar populations are more promising 
candidates. The Peimbert Class \citep{peim78, fm87} gives an estimate of 
the progenitor age (as well as abundance criteria similar to ours). Our 
targets are chosen from classes IIb, III, and IV. 

\item Since we observe \hep3, our target PNe must have excitations such
that most of the He is singly ionized.

\end{itemize}

Here we discuss \hep3\ VLA observations of two PNe: \object[NGC
6572]{\ngc{6572}} and \object[Jonckheere 320]{\jon{320}}. Both objects
are expected to have high \her3\ abundance ratios; they were selected
using criteria similar to those above.  In addition, the angular sizes
of these PNe are well matched to the VLA synthesized beam at 8.7\ghz.

\section{Observations}

\begin{figure}
\includegraphics[angle=-90, scale=0.7]{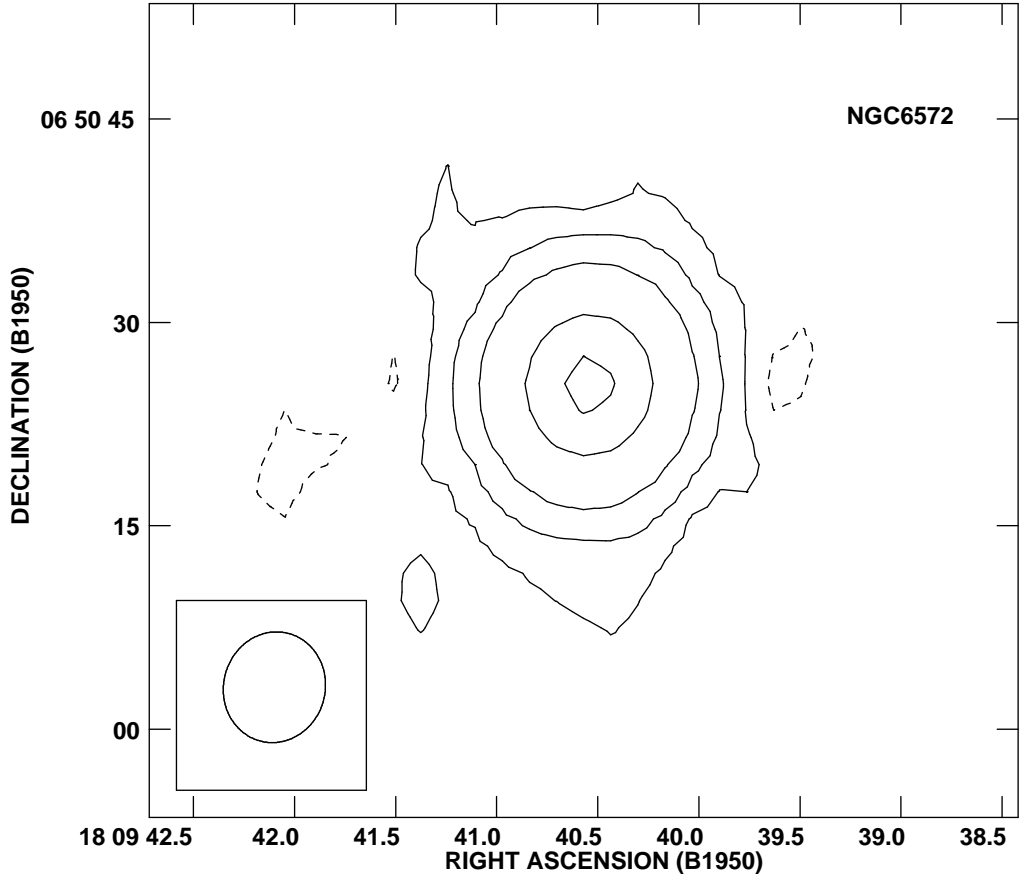}
\figcaption{
VLA continuum image of \ngc{6572} at 8.7$\,$GHz.  The resolution, 
$8\arcsper 20\times 7\arcsper 49$ at a PA of $-14\,\degr\,$, 
is indicated in the lower left-hand corner.  The intensity scale is 
in units of mJy$\,$beam$^{-1}$.  The contour levels are at $-$3, 3, 
25, 100, 500, and 1000 times the $3\,\sigma$ (0.834\mjy) noise level.
\label{fig:n6572_cont}}
\end{figure}

\begin{figure}
\includegraphics[angle=-90, scale=0.7]{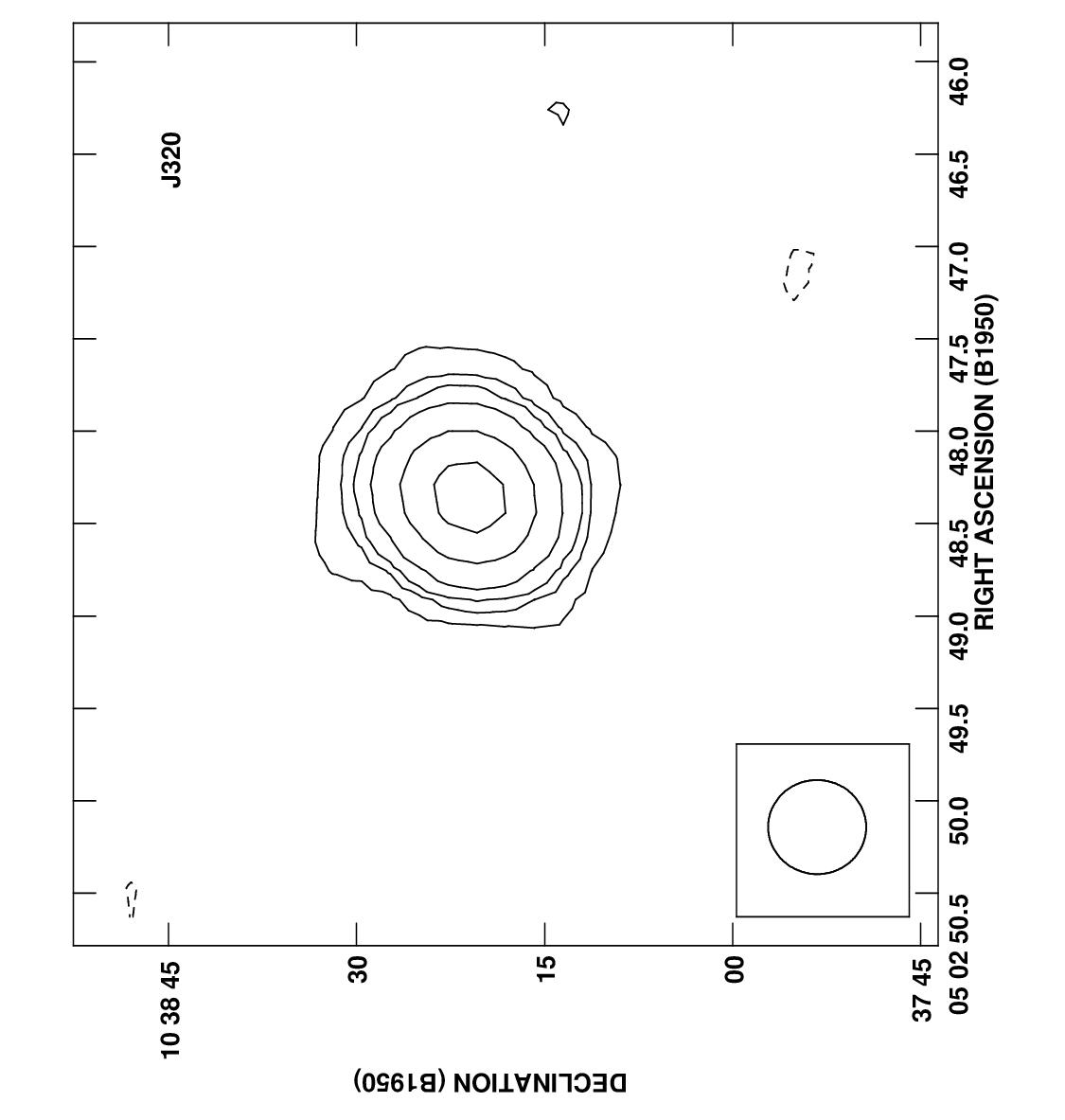}
\figcaption{
VLA continuum image of \jon{320} at 8.7$\,$GHz.  The resolution,
$7\arcsper 82\times 7\arcsper 51$ at a PA of $-2\,\degr\,$, 
is indicated in the lower left-hand corner.  The intensity scale is 
in units of mJy$\,$beam$^{-1}$.  The contour levels are at $-$1, 1, 
3, 5, 10, 25, and 50 times the $3\,\sigma$ (0.264\mjy) noise level.
\label{fig:j320_cont}}
\end{figure}

\placetable{tab:vla}
\placefigure{fig:n6572_cont}
\placefigure{fig:j320_cont}

Table~\ref{tab:vla} summarizes the VLA\footnotemark[1] D-array configuration 
observations.
\ngc{6572} was observed on five different days.  The correlator spectrometer 
was setup to observe the \hep3\ hyperfine transition at two orthogonal 
polarizations at a rest frequency of 8665.65\mhz.  The data were Hanning 
smoothed on-line.  During part of the first day the \hal\ transition was 
observed at a rest frequency of 8584.81\mhz.  \jon{320} was observed on 
seven different days.  The correlator was setup to observe the \hep3\ and 
\hal\ transitions simultaneously at two orthogonal polarizations each.  
The data were Hanning smoothed off-line.

\footnotetext[1]{\nraoblurb}

The data were edited and calibrated using the Astronomical Image
Processing System (AIPS).  The calibrator 3C\,286 was used to
calibrate the intensity scale.  Because the \hep3\ emission line
intensity was expected to be weak the bandpass had to be carefully
calibrated.  The spectral baselines are typically much flatter for
interferometers than single-dish telescopes since any telescope
dependent effect will be removed when the data from different antennas
are cross correlated.  But at the sensitivity levels required for our
observations the VLA bandpass is not stable with time; amplitude
variations as large as 1\% have been measured with an approximate
width of 3\mhz.\footnotemark[2] These baseline ripples are caused by
reflections in the waveguides and move across the bandpass at about
0.4\mhz\ per hour.  We therefore observed a bandpass calibration
source approximately every 15 minutes for a duration of about 5
minutes.  A bandpass model was calculated by interpolating the
bandpass solutions over time.  We are confident that we have removed
most of this structure since the measured spectral rms noise is
consistent with the expected thermal noise.

\footnotetext[2]{See ``A Guide for Spectral Line Observers'' 
at http://www.vla.nrao.edu/astro/guides/sline/current/}

The continuum intensities of \ngc{6572} and \jon{320} were sufficient
to perform the standard self-calibration technique.  We applied the
self-calibration algorithm to each day's observations after the
standard phase and bandpass calibration.  This procedure typically
improved the signal-to-noise ratio by a factor of five in the
continuum.  The task DBCON was used to average the data.  The
self-calibration technique was applied to the final average.  The task
UVLSF was used to subtract the continuum emission in the ($u,v$)-plane
using the line-free channels.  This fits a linear baseline to the
specified channels in the real and imaginary parts of the visibilities.
Continuum and line images were produced using the CLEAN algorithm.
Table~\ref{tab:vla} lists the rms noise in these images. The continuum
images are completely limited by the finite dynamic range of the
strong continuum and thus are not limited by thermal fluctuations. The
line data, after subtraction of the continuum, are limited by thermal
noise. Thus the continuum rms values are larger than the line data.

\placetable{tab:cont}

\section{Observational Results}

The continuum images of \ngc{6572} and \jon{320} are shown in
Figures~\ref{fig:n6572_cont} and \ref{fig:j320_cont}, respectively.
The continuum emission is well modeled by a Gaussian function.  The
task JMFIT was used to fit a single-component, two-dimensional
Gaussian model to the continuum image for each PN.
Table~\ref{tab:cont} summarizes the results.  Listed are the source
name, the B1950 equatorial coordinates of the peak radio emission, the
peak flux density, the integrated flux density, and the angular size.

We analyzed the \hal\ and \hep3\ spectral line data cubes using four
different methods: (1) the ISPEC task was used to produce spectra
toward the continuum peak emission within a single resolution element;
(2) averages over different spatial areas were made in order to
increase the signal-to-noise ratio; (3) the images were convolved with
Gaussian functions that had half-power beam-widths (HPBWs) larger than
the synthesized beam; and (4) the task IRING was used to compute averages
over concentric annuli to produce a cumulative flux density versus
radius relation.  This latter method averages over both position and
velocity.

\begin{figure}
\includegraphics[angle=-90, scale=0.7]{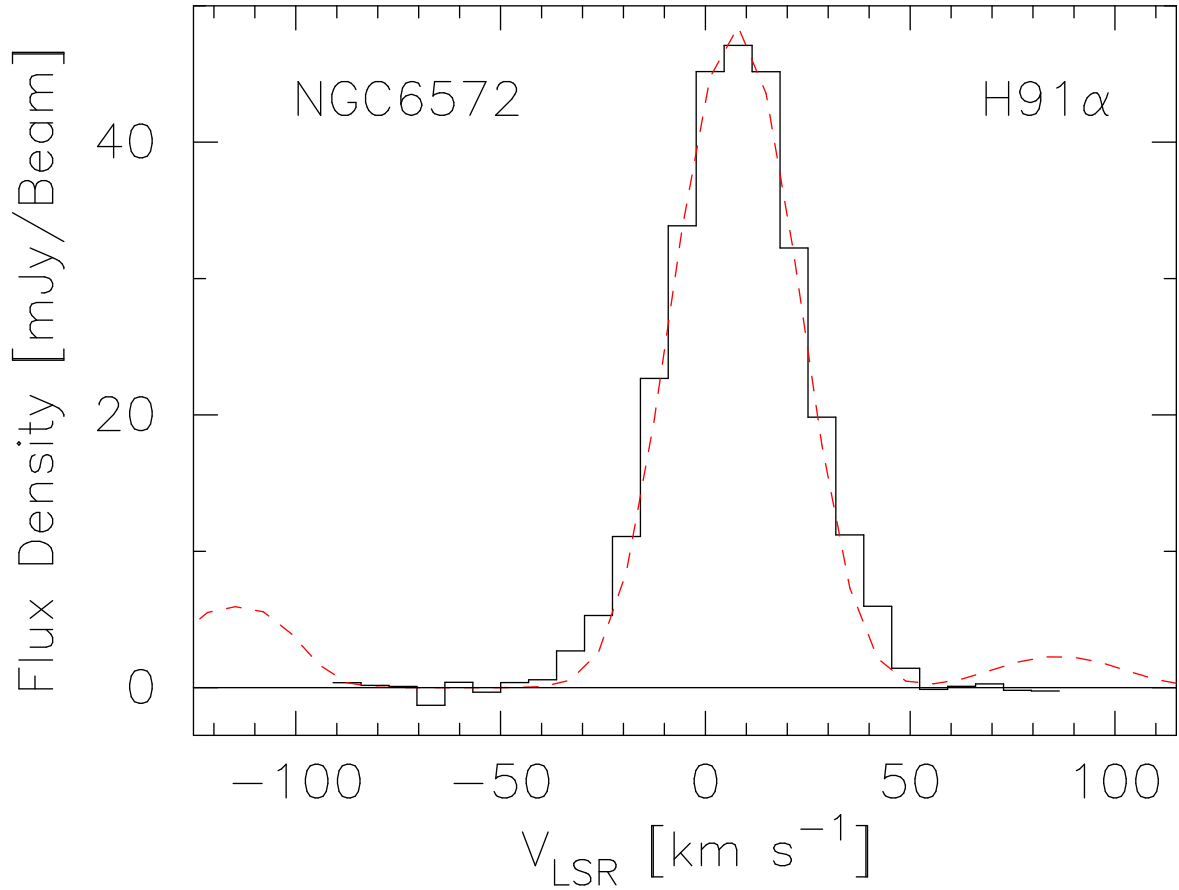}
\figcaption{\hal\ spectrum of \ngc{6572}.  The solid histogram is the
VLA data toward the peak continuum position.  The dashed curve shows 
the NEBULA model assuming  $V_{\rm LSR} = 8$\kms.
The modeled line at negative velocities is the \heal\ transition whereas 
the modeled line at positive velocities is the \heps\ transition.
The rms over the line-free spectral regions is 500\microjyb.
\label{fig:n6572_hal_model}}
\end{figure}

\begin{figure}
\includegraphics[angle=-90, scale=0.7]{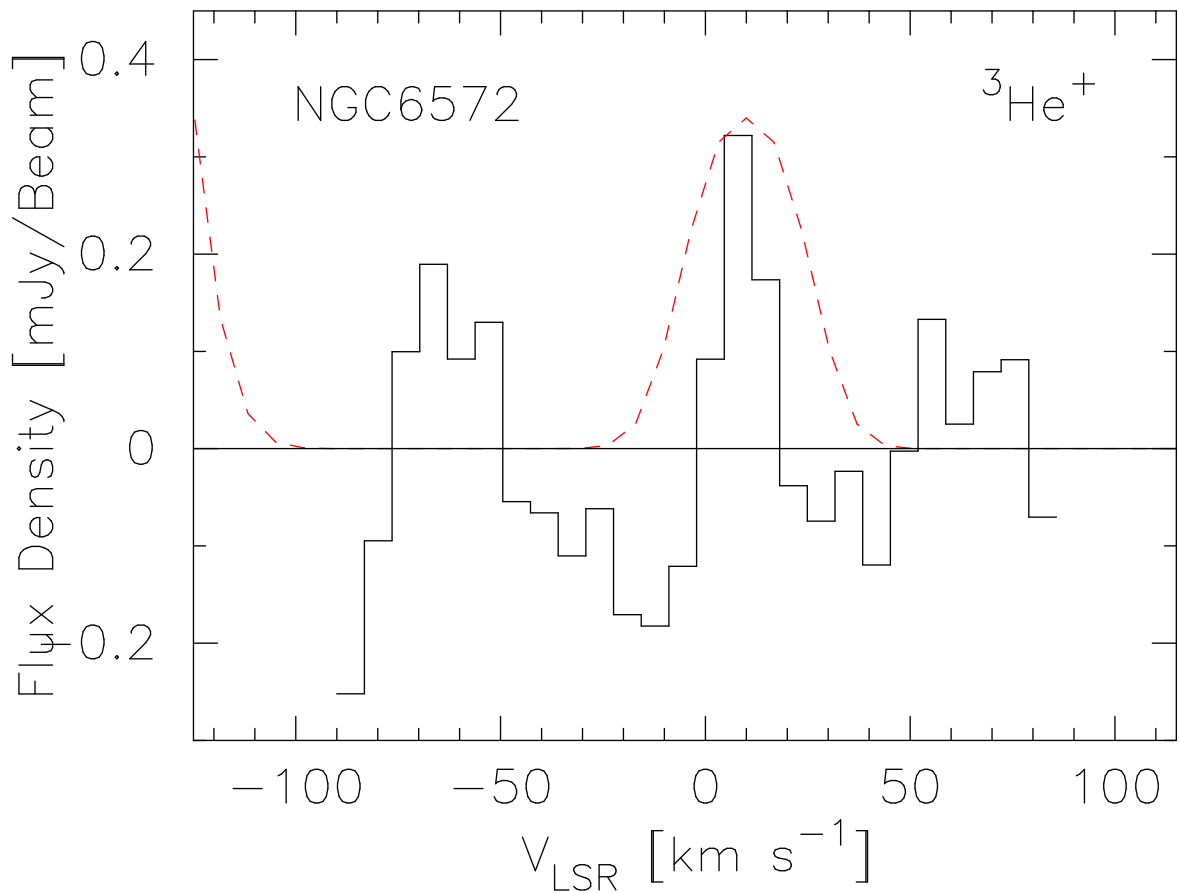}
\figcaption{\hep3\ spectrum of \ngc{6572}.  The solid histogram is the
VLA data toward the peak continuum position.  The dashed curve shows
the NEBULA model assuming $V_{\rm LSR} = 8$\kms.  The modeled line at
negative velocity is the \heta\ transition.  The rms over the
line-free spectral regions is 100\microjyb. The model \hepr3\
abundance ratio is \nexpo{1.0}{-3} by number.
\label{fig:n6572_he3_model}}
\end{figure}

\begin{figure}
\includegraphics[angle=-90, scale=0.65]{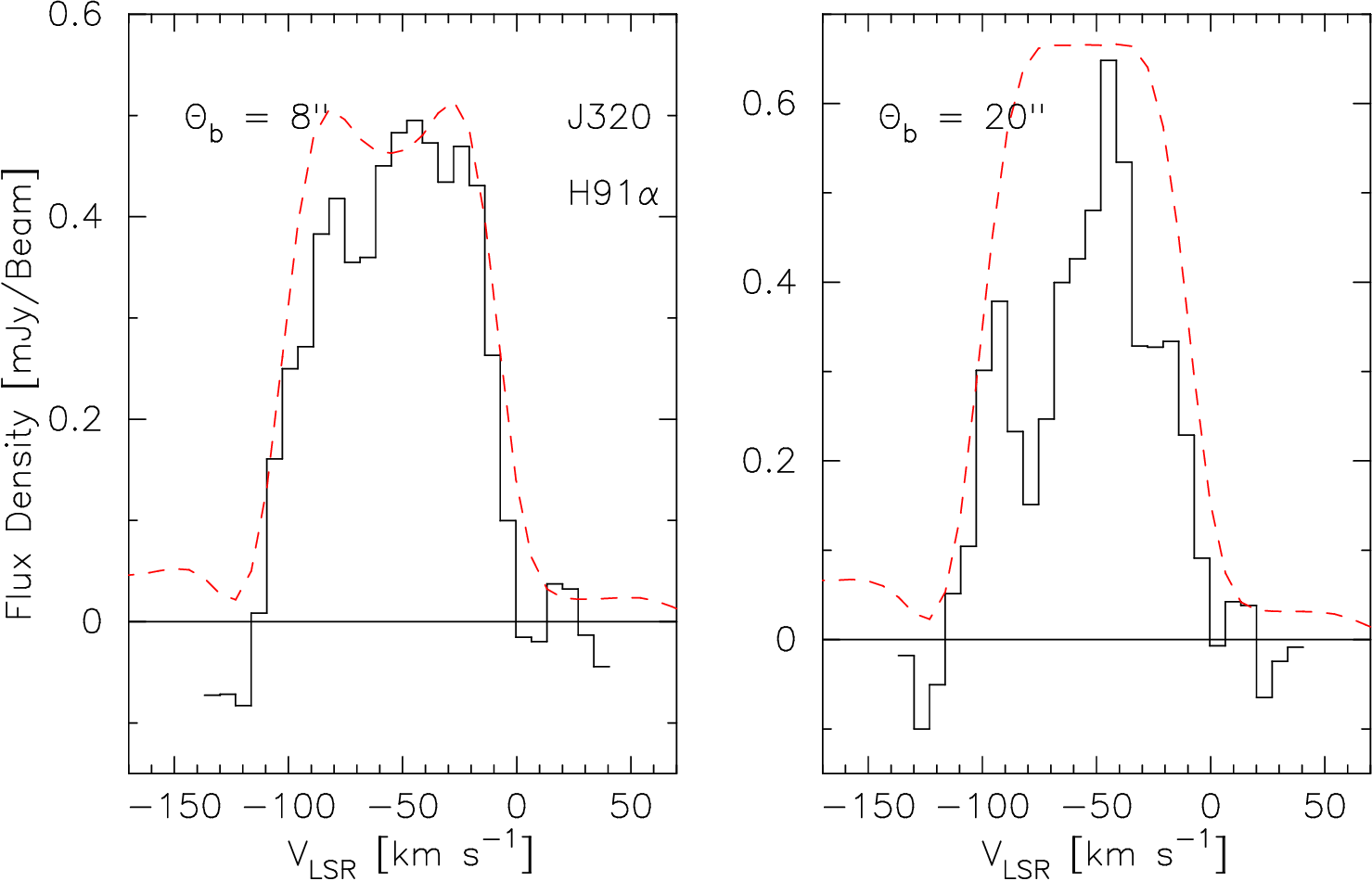}
\figcaption{\hal\ spectra of \jon{320}.  The solid histograms are
the VLA data and the dashed curves are the NEBULA model results
assuming $V_{\rm LSR} = -54.9$\kms.  Left panel: spectrum toward
the peak continuum position for a synthesized HPBW of 8\arcsec.  The
rms over the line-free spectral regions is 50\microjyb.  Right panel:
spectrum toward the peak continuum position for a synthesized HPBW of
20\arcsec.  The VLA image was convolved with a circular 20\arcsec\ beam
using the task CONVL.  The rms over the line-free spectral regions is
70\microjyb.
\label{fig:j320_hal_model}}
\end{figure}

\begin{figure}
\includegraphics[angle=-90, scale=0.65]{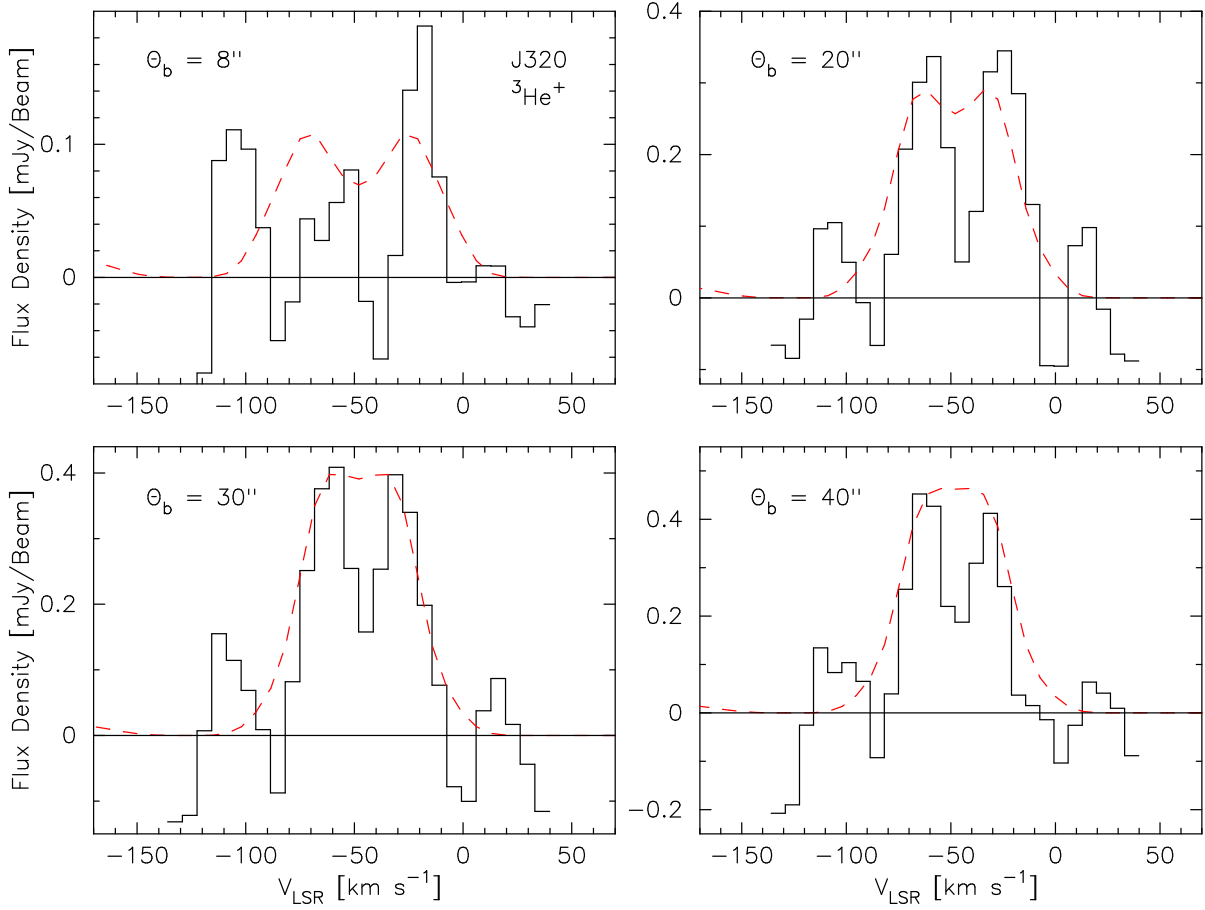}
\figcaption{\hep3\ spectra of \jon{320}.  The solid histograms are
the VLA data and the dashed curves are the NEBULA model results
assuming $V_{\rm LSR} = -47.9$\kms.
The VLA data were convolved with a circular beam using the task
CONVL.  The HPBW is shown in the top left corner of each panel.  The
rms values over the line-free spectral regions are 70, 110, 150, and
190\microjyb\ for HPBW's of 8, 20, 30, and 40\arcsec, respectively. In
the latter three cases the estimated $S/N$ is 4.
The model \hepr3\ abundance ratio is \nexpo{1.9}{-3} by number.
\label{fig:j320_he3_model}}
\end{figure}

\begin{figure}
\includegraphics[angle=-90, scale=0.7]{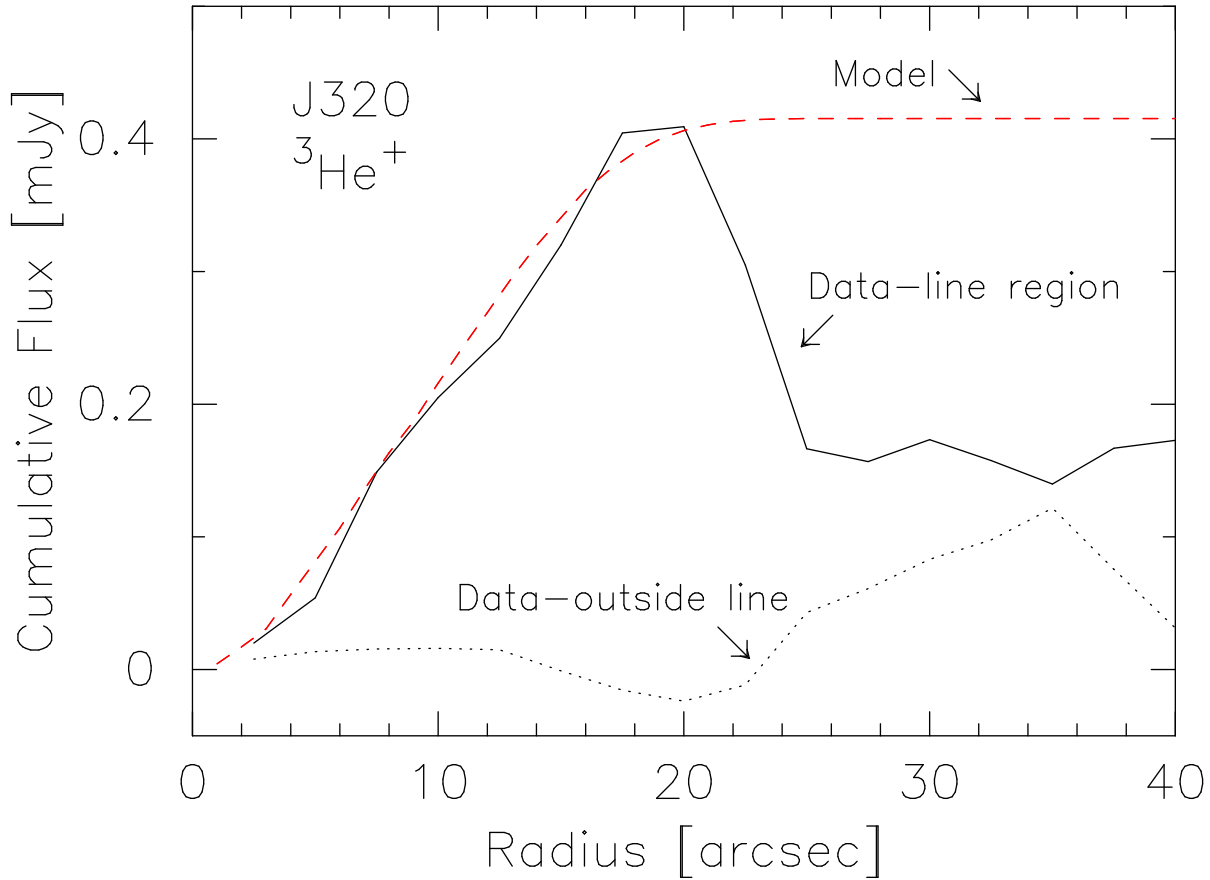}
\figcaption{Cumulative \hep3\ flux density versus radius for
\jon{320}.  The task IRING calculates the integrated flux density in
concentric annuli from a specified center location.  The solid curve
is the cumulative flux density for the \hep3\ image using the observed
peak continuum emission as the center location. The cumulative flux at
larger radii drops because of the missing short spacings. The dashed
lines are the result of a similar analysis using the modeled data
assuming a halo size of 35\arcsec\ and an electron density of 50\percc.
The model \hepr3\ abundance ratio is \nexpo{1.9}{-3} by number. The
missing short spacings do not affect the model.  The
dotted curve shows the result of a similar analysis made for \hep3\
emission-free data channels.  The rms of the cumulative flux in
this noise image within a radius of 20\arcsec\ is 43\microjy\ whereas 
the peak cumulative \hep3\ flux
is 400\microjy\ yielding a $S/N$ for the line of $\sim 9$.
\label{fig:j320_iring_abund}}
\end{figure}

\placefigure{fig:n6572_hal_model}
\placefigure{fig:n6572_he3_model}
\placefigure{fig:j320_hal_model}
\placefigure{fig:j320_he3_model}
\placefigure{fig:j320_iring_abund}

We detected \hal\ line emission in both PN; only \jon{320} has \hep3\
emission.  Spectra are shown in Figures~\ref{fig:n6572_hal_model} and
\ref{fig:n6572_he3_model} for \ngc{6572} and in
Figures~\ref{fig:j320_hal_model} and \ref{fig:j320_he3_model} for
\jon{320}.  The solid line histograms are the data and the dashed
lines are models described in \S\,{\ref{sec:models}}.  The additional
spectra shown for \jon{320} were made by convolving the \hal\ data
cube with a HPBW of 20\arcsec\ (Figure~\ref{fig:j320_hal_model}) and
the \hep3\ data cube with HPBW's of 20, 30, and 40\arcsec\
(Figure~\ref{fig:j320_he3_model}). Comparing the \hep3\ line intensity 
to the noise level in the line free region in Figure~\ref{fig:j320_he3_model} 
shows that the detection has a signal to noise ratio of $S/N \sim 4$.  
Figure~\ref{fig:j320_iring_abund} plots the cumulative \hep3\ flux density 
from the \hep3\ channels (12--22) versus radius for \jon{320}.  The \hep3\ 
flux density gradually increases until a radius of 15--20\arcsec.  For 
larger radii the flux drops because of missing short spacing data. In the 
configuration used for these observations the VLA has no sensitivity to 
\hep3\ emission with radii larger than about 30\arcsec. The plot also shows 
the cumulative flux density for the channels not containing the line 
emission (6--11 \& 23--26). A comparison of the line and line-free results 
gives a $S/N$ of $\sim 9$ for the \hep3\ detection. This value is larger than 
that obtained from the spectra because the \hep3\ emission is averaged over 
both space and frequency. The detected continuum and recombination line emission 
are both confined to a smaller area within a radius of 5--10\arcsec.  Apparently 
much of the \hep3\ emission from \jon{320}\ emanates from a large, low density
halo. The model shown in Figure~\ref{fig:j320_iring_abund} has a 35\arcsec\ halo 
(see \S\,\ref{sec:j320}). This halo model fits the data well except for larger 
radii where it is not affected by the missing short baselines.  

\section{The ${\bf {}^3He}$ Abundance}\label{sec:models}

Estimates for the \her3\ abundance ratios in the PNe were made by 
numerical modeling.  
We used the radiative transfer code NEBULA to calculate synthetic
spectra of the radio continuum, radio recombination line, and 
\hep3\ line emission from a model nebula \citep{balser99}.  The
model is comprised solely of ionized hydrogen and helium gas.  A
density and ionization structure is specified within a three-dimensional 
Cartesian coordinate system.  For each numerical cell the following 
physical properties are specified: electron temperature ($T_{e}$), 
electron density ($n_{e}$), and helium ionization (\hepr4, \heppr4, 
\hepr3).  The modeled nebula consists of homogeneous concentric shells 
which have a radial expansion velocity \citep{balser97, balser99b}. The 
line profile is therefore broadened by three components: thermal, 
microturbulent, and large-scale expansion. The thermal motions are determined 
from the electron temperature, whereas the microturbulent motions, by 
definition smaller than the numerical cell, are set by a single free 
parameter that is constrained by the observed linewidths. The \hep3\ 
and recombination lines are assumed to be in LTE.  Non-LTE effects and
pressure broadening from electron impacts are negligible for these PN models.
Synthetic 
spectra are generated by calculating the radiative transfer for the line 
and continuum emission as a function of frequency from the back to the 
front of the numerical grid.  The NEBULA outputs are the two-dimensional 
continuum brightness distribution and spectral data cubes for \hep3\ and 
the recombination lines.
These model data are then convolved with a specified telescope beam so
they can be compared directly with observations.   

We use an iterative procedure to derive an estimate for the \her3\ abundance.
First the physical properties of a source are constrained by the continuum and 
recombination line emission as well as data from the literature.  A fiducial
\her3\ abundance is then chosen and synthetic spectra are calculated with 
NEBULA.  The \her3\ abundance is iterated until the model spectra match
the observations.  A grid of models is used to assess the uncertainty in
the derived \her3\ abundance.  

\placetable{tab:model}

The planetary nebulae properties used as input for NEBULA are
summarized in Table~\ref{tab:model}.  The PNe distance, expansion
velocity, and helium ionization structure are taken from the
literature (see \S\,{\ref{sec:n6572}--\ref{sec:j320}}).  The VLA data
are used to model the PNe as single, homogeneous, spheres
\citep{balser95, balser99}.  From these models we determine the
spherical angular size, electron temperature, and electron density.
Below we discuss each PN separately.

\subsection{\ngc{6572}}\label{sec:n6572}

\ngc{6572}, discovered by Struve in 1825, is a well studied 
planetary nebula \citep{acker92}.  Unlike most PNe the distance to
\ngc{6572} has been accurately measured by combining the angular
expansion rate, determined from high spatial resolution radio
continuum images at two or more epochs, with the Doppler expansion
velocity \citep{masson86}.  We adopt a distance to \ngc{6572} of 1.2\kpc\
\citep{kawamura96}.  The expansion velocity is directly measured from
spectral line data that resolves the PN shell \citep{masson89,
acker92}.  The helium ionization properties are taken from
\citet{cahn92}.

The VLA data were used to model a single, homogeneous sphere with an
angular size of $7\farcs 9$.  Although \ngc{6572} is just resolved by
our VLA data this size is consistent with other estimates
\citep{acker92}.  \ngc{6572} has been imaged at higher resolution in
the radio \citep{masson89}, near infrared \citep{hora90}, and optical
\citep{miranda99}.  Morphologically, \ngc{6572} has a double-lobed,
elliptical structure with a central minimum.  We assume an inner shell
size of 2\arcsec.  The electron temperature and density are found
to be 10,300 K and 22,000\percc, respectively.  Various values for
$T_{e}$ and $n_{e}$ can be found in the literature. These values
depend on the spectral lines employed, the method used to determine
these physical parameters, and the volume within the nebula sampled
since real temperature and density fluctuations are present
\citep{mathis98}.  For example, electron temperatures determined from
optical forbidden line ratios are weighted toward higher temperature
regions whereas radio recombination line-to-continuum ratios are
weighted toward lower temperature regions \citep{mansfield69}.
Nevertheless, our estimates are in good agreement with
\citet{zhang04}, who use the optical hydrogen recombination spectrum
to determine $T_{e}$ and $n_{e}$.

The model results are shown as dashed curves in
Figures~\ref{fig:n6572_hal_model} and \ref{fig:n6572_he3_model} for
the \hal\ and \hep3\ lines, respectively.  The \hal\ model profile
is a good fit to the observations.  The data are consistent with
an upper limit for the \her3\ abundance ratio of $\la$ \expo{-3}.

\subsection{\jon{320}}\label{sec:j320}

Discovered in 1913 \citep{jonckheere13}, \jon{320} has a radio continuum 
flux density about 50 times weaker than \ngc{6572};  thus it is not as 
well studied.  The morphology and kinematics of \jon{320} have recently 
been investigated using long slit spectroscopy with the Anglo-Australian 
Telescope (AAT) and \halpha\ images with the Hubble Space Telescope (HST)
\citep{harman04}.  The central nebula consists of two (possibly three)
bipolar lobes expanding at $\sim 46$\kms\ with different orientations.
The morphology is thus better classified as a poly-polar PN.
Surrounding the central nebula are high speed knots that may not have
played a significant role in forming the central nebula.

The distance estimates to \jon{320} vary from 2 to 6\kpc\ using
different methods \citep[and references within]{harman04}.  Here we
adopt a distance of 5\kpc.  
The helium ionization properties are taken from \cite{cahn92}.  The
VLA radio continuum and recombination line data are used to determine
the angular size, electron temperature, and electron density. The
angular size is $7\farcs 4$, consistent with previous estimates
\citep{acker92, tylenda03}.  We assume an inner shell size of 2\arcsec\,. 
The electron temperature and density are 7,500 K and
1,500\percc, respectively; these values are lower than predicted by
optical forbidden line ratios \citep{barker78, cahn92}, as might be
expected for a nebula with temperature and density fluctuations.

Because the \hep3\ emission extends beyond the continuum emission, we
infer that a low density halo exists for this PN (see
Figure~\ref{fig:j320_iring_abund}).  Halo components are not uncommon
for PNe but are difficult to detect in optical (H$\alpha$ emission) or
radio continuum (free-free emission) images since these tracers of
ionized gas are proportional to the emission measure ($\propto n_{\rm
e}^2$) and halo densities are typically low.  The \hep3\ emission,
however, is sensitive to the total column density ($\propto n_{\rm
e}$).  Therefore it is plausible to detect the halo in \hep3\ emission
but not radio continuum emission. The same situation obtained
for \ngc{3242} \citep{balser97, balser99b}.

\begin{figure}
\includegraphics[angle=-90, scale=0.7]{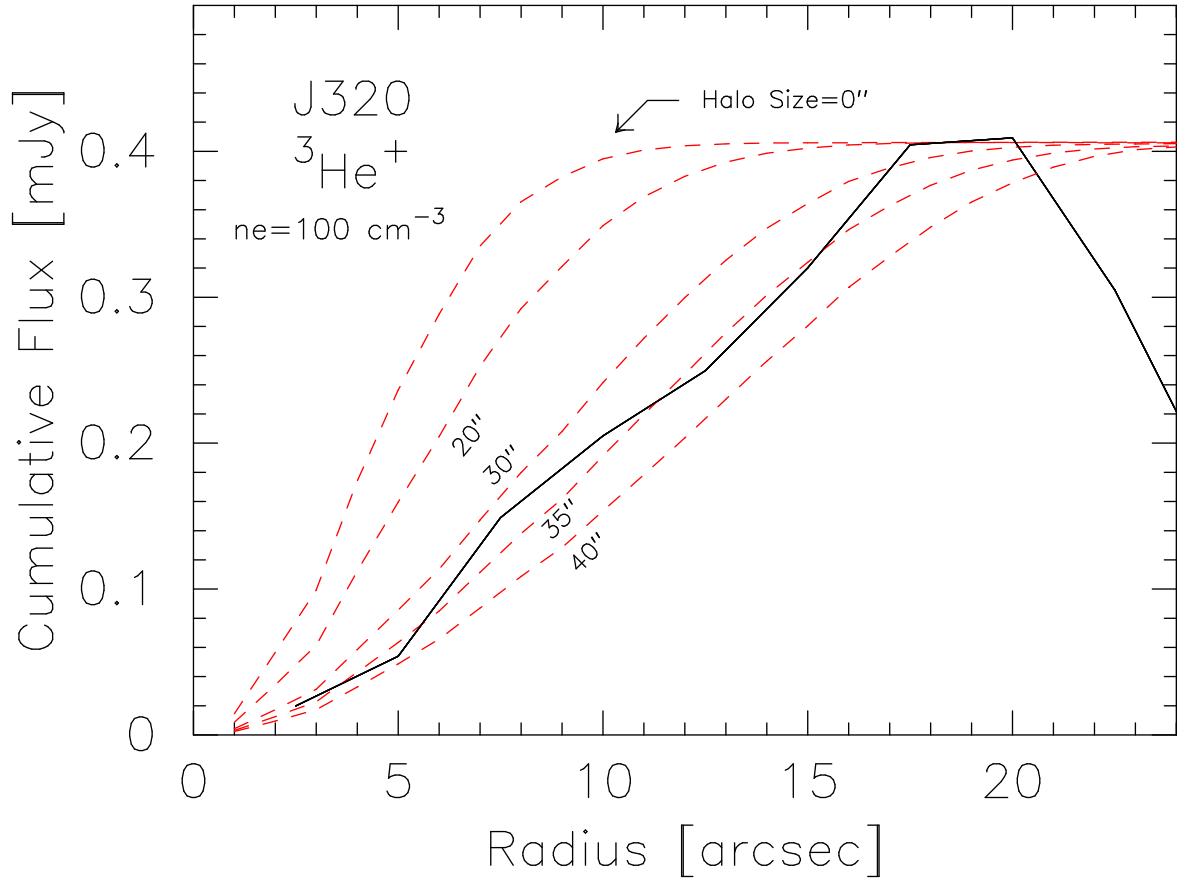}
\figcaption{
Cumulative \hep3\ flux density versus radius for \jon{320}.  The solid curve is 
that of Figure\,\ref{fig:j320_iring_abund}.  The dashed lines are the results of a 
similar analysis using modeled data.  The models have a 
halo electron density of 100\percc\ and different halo sizes from left
to right: 0, 20, 30, 35, and 40\arcsec.  The model results were 
scaled to match the peak \hep3\ cumulative flux density.
\label{fig:j320_iring_ne}}
\end{figure}

\begin{figure}
\includegraphics[angle=-90, scale=0.7]{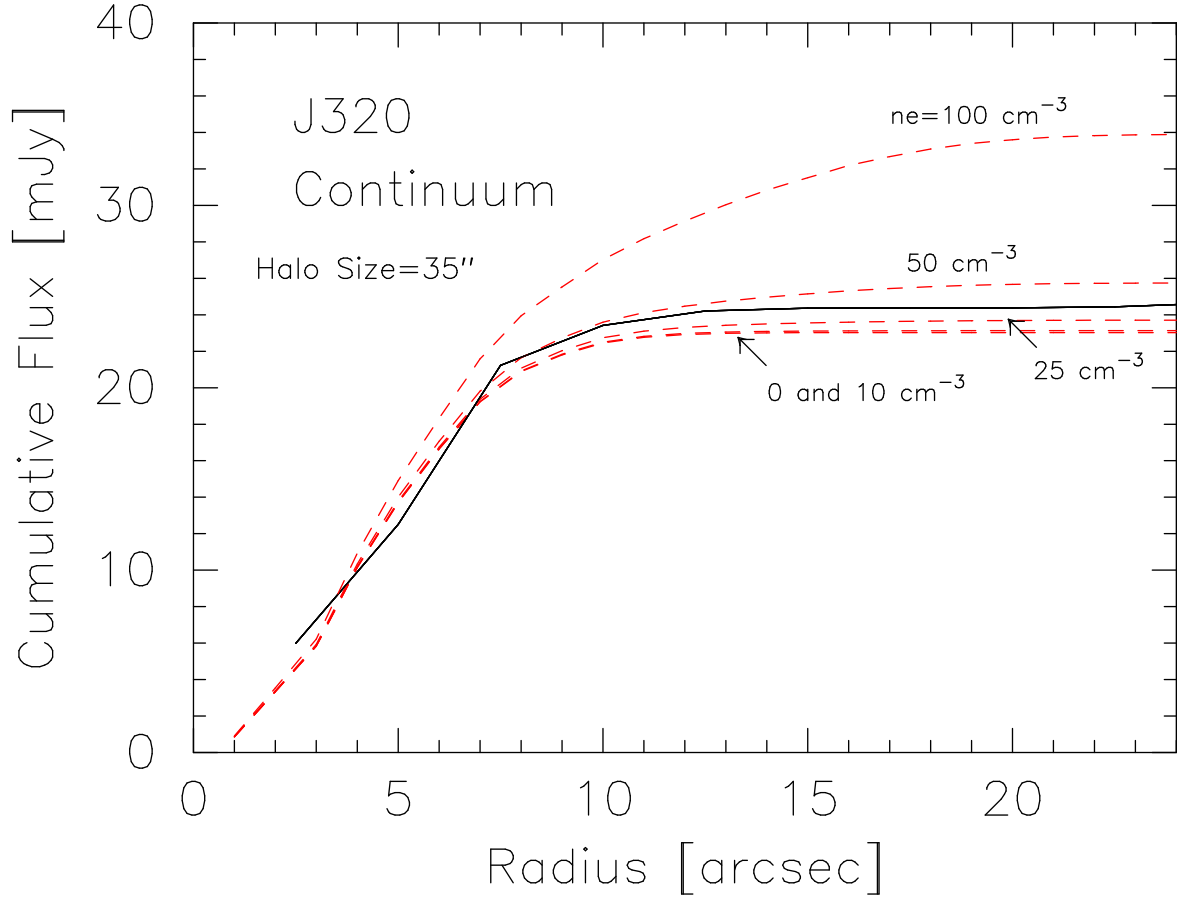}
\figcaption{
Cumulative continuum flux density versus radius for \jon{320}.  
The solid curve is the cumulative flux density for the continuum image 
using the observed peak continuum emission as the center location.  
The dashed lines show the result of a similar analysis using modeled 
data.  The models have a halo size of 35\arcsec and different halo electron 
densities from bottom to top: 0, 10, 25, 50, and 100\percc.
\label{fig:j320_iring_halo}}
\end{figure}

\placefigure{fig:j320_iring_ne}
\placefigure{fig:j320_iring_halo}

The \hep3\ and radio continuum data together with NEBULA models 
were used to constrain some of the halo properties.  To estimate the size
of the halo we modeled the \hep3\ emission using NEBULA for
different halo sizes.  Figure~\ref{fig:j320_iring_ne} plots the
cumulative \hep3\ flux density versus radius for the VLA data (solid
curve) and for five different halo models (dashed curves).
The electron density was assumed to be 100\percc\ for all halo
models and the cumulative \hep3\ flux density was scaled to match the VLA
data since we are only interested in the relative shape.
Qualitatively a halo size of 35\arcsec\ best fits the data.  
We set a limit for the halo electron density by calculating the
continuum emission for different values of $n_{e}$.
Figure~\ref{fig:j320_iring_halo} plots the cumulative continuum flux
density versus radius for the VLA data (solid curve) and for five
models (dashed curves) with different halo electron densities and a
halo size of 35\arcsec.  For halo densities between
$0<n_{e}<50$\percc\ the predicted continuum emission is within 5\% of
the measured VLA continuum.  The upper limit of $n_{e} = 50$\percc\
sets a lower limit to the predicted \her3\ abundance ratio.  The
expansion velocity of the halo is constrained by the \hep3\ profile
shape.  Given the low signal-to-noise ratio of the \hep3\ line and the
complex morphology of \jon{320}, it is not clear how to interpret the
double profile in Figure~\ref{fig:j320_he3_model}.  In our models
the Doppler thermal and microturbulent broadening together produce a 
Gaussian line shape.  An expansion broadening, however, produces either 
a square-wave profile for an unresolved source or two separate peaks 
for an optically thin, resolved source.  Nevertheless, in the context of 
our spherically symmetric expanding shell model we use the overall width 
of the \hep3\ profile to estimate a halo expansion velocity of 25\kms.

Based on these constraints, we adopt a halo size of 35\arcsec\,, an
electron density of 50\percc\,, and a uniform expansion velocity of
25\kms.  Because we have no information about these physical
properties in the halo, the electron temperature and ionization
properties in the halo are assumed to be the same as the core values.
Since the \her3\ abundance ratio is only weakly dependent on the
electron temperature, having to use an assumed value is not a serious
compromise.  Furthermore, since the source of ionization is a hot
central star all of the helium should be singly ionized.

The results are summarized in
Figures~\ref{fig:j320_hal_model}--\ref{fig:j320_iring_abund}.  The
solid lines are the data whereas the dashed curves are the model results 
assuming a \her3\ abundance of \nexpo{1.9}{-3}.  Overall the model fits 
the data reasonably well.  
The major uncertainties in determining the \her3\ abundance ratio in
\jon{320} are: (1) the measured line intensity; (2) the distance; 
(3) the halo electron density; and (4) the halo size.  The \her3\
abundance ratio is proportional to $T_{\rm L}\,R_{\rm sun}^{-1/2}$,
where $T_{\rm L}$ is the \hep3\ line intensity and $R_{\rm sun}$
is the distance \citep{rood84}.  Depending on how the signal-to-noise
ratio is estimated, we detect \hep3\ emission at a S/N of $\sim\,$ 4
or 9. Taking the more conservative value of 4, the uncertainty in the
\her3\ abundance is \nexpo{\pm\ 0.5}{-3}.  An error of
1.5\kpc\ in the distance gives an error in the abundance of
\nexpo{\pm\ 0.3}{-3}.  The VLA continuum image limits the halo
electron density to less than 50\percc.  Smaller values of $n_{e}$ are
possible and will increase \her3.  For example, a halo electron
density of 10\percc\ will increase \her3\ by 50\%.  The halo size is
constrained by the distribution of \hep3\ emission.  But a larger halo
may exist and not be detected by our observations.  For example, if
the halo size extends beyond about 60\arcsec\ it would not be detected in
\hep3\ emission with the VLA. A larger halo will reduce our predicted
value of \her3.  For example, a halo size of 100\arcsec\ will decrease
the \her3\ abundance by 50\%. None of these errors is likely to be
normally distributed, so we cannot combine them in a meaningful way.
Crudely, we estimate that the abundance error is roughly a factor of
two.

\section{Discussion}

The most important result of this paper is that we have detected
\hep3\ in another planetary nebula, \jon{320}, with a \her3\ 
abundance ratio roughly {\it two orders of magnitude larger} than
the primordial value (\her3\,$\sim$\,\expo{-5} by number) determined
from Galactic \hii \ region abundances and confirmed by WMAP cosmic
microwave background results.  Merely detecting \hep3\ in \jon{320}
tells us that its stellar progenitor evolved in a manner well
described by standard stellar models: it produced \he3\ during its
main sequence lifetime; it did not undergo any extra-mixing; it
subsequently expelled \he3\ enriched gas into the ISM.
To be consistent with chemical evolution models, the existence of the
\he3\ Plateau demands that the majority of PNe are not net producers of \he3.
\jon{320} now joins \ngc{3242} as only the second known example of
those rare planetary nebulae which enrich the Galaxy in \he3.

Our long term goal is to identify some surrogate characteristics which
allow us to determine what fraction of planetary nebulae enrich the ISM
in \he3\ without having to measure \he3\ itself. Now that the sample of 
\he3\ PNe has a population of two we can begin this attempt.  There are some
intriguing similarities between \ngc{3242} and \jon{320}. They both
satisfy our selection criteria: each has low He/H and [N/O] abundance 
which indicates little or no extra-mixing; each has most of its
He in a singly ionized state; each originates from an old stellar
population as indicated by its Peimbert Class (\ngc{3242}
is IIb; \jon{320} is IV). 

Superficially, all Peimbert Class IIb, III \& IV PNe should be good
\he3\ targets. Class IIb alone encompasses about 35\% of all the known
PNe (Quireza, private communication). In order to not over-produce \he3, 
there must be something in addition to either the Peimbert Class or our 
other selection criteria (or both) that determines which PNe have high 
\he3\ abundances. With the detection of \he3\ in \jon{320} we
have a first hint as to what that parameter might be. Both detected
PNe have low density halos which apparently produce the bulk of the
\hep3\ signal that we detect \citep[and herein]{balser99b}.  Both have
lineshapes that are consistent with optically thin expanding
shells. Are low density halos required for high \he3, or do they
simply make it easier to detect if present?

\section{Conclusions}

We detected \hep3\ emission with the VLA in the planetary nebulae
\jon{320}.  Using numerical models we derive a \her3\ abundance ratio
by number of \nexpo{1.9}{-3} with an uncertainty of roughly a factor
of two.  This value is two orders of magnitude higher than the typical
\he3\ abundance found in Galactic \hii\ regions.
This detection makes \jon{320} the second planetary nebula known 
(together with \ngc{3242}) to have an anomalously high \he3\ abundance.

\acknowledgements The research was supported by the National Science
Foundation (AST 00-98047). We thank Cintia Quireza for discussions
about PNe and providing her
PNe database prior to publication. We thank the international light
elements community for their friendship and support.

{\it Facilities:} \facility{VLA}

\newpage


\newpage



\begin{deluxetable}{lll}
\small
\tablecolumns{3}
\tablecaption{VLA Observational Parameters}
\tablehead{
\colhead{Parameters} & \colhead{\ngc{6572}} &
\colhead{\jon320}
}
\startdata
Date                            & 1995 February                       & 1996 August  \\
Total Time (hr)$^{\rm a}$       & 24                                  & 23  \\
Configuration                   & D                                   & D \\
R.A. of field center (B1950)    & $18^{\rm h}09^{\rm m}40.57^{\rm s}$ & $05^{\rm h}02^{\rm m}48.6^{\rm s}$ \\
Dec. of field center (B1950)    & $06\degr\,50\arcmin\,25\arcsper 5$  & $10\degr\,38\arcmin\,25\arcsper 0$ \\
FWHM of primary beam            & $5\arcmper 4$                       & $5\arcmper 4$ \\
FWHM of synthesised beam        & $8\arcsper 20\times 7\arcsper 49$   & $7\arcsper 82\times 7\arcsper 51$ \\
PA of beam                      & $-14\degr$                     & $-2\degr$ \\
 & & \\
Total bandwidth (MHz)           & 6.25                                & 6.25 \\
Number of channels              & 31                                  & 31 \\
LSR central velocity (\kms)     & 8.0                                 & --37.9  \\
Spectral resolution (kHz, \kms) & 195.3 (6.8)$^{\rm b}$               & 326.2 (11.4)$^{\rm c}$ \\
 & & \\
Flux density calibrator         & B1328+307/B0134+329                 & B1328+307/B0134+329 \\
Phase calibrator                & B1730--130/B1749+096                & B0446+112/B0316+413 \\
 & & \\
Line channel $RMS$ (\microjyb)$^{\rm d}$    & 100 & 70 \\
Continuum $RMS$ (\microjyb)$^{\rm d,e}$ & 280 & 90  \\

\enddata
\tablenotetext{a}{Approximate time not excluding edited data.}
\tablenotetext{b}{On-line Hanning smoothed.}
\tablenotetext{c}{Off-line Hanning smoothed.}
\tablenotetext{d}{Using the \hep3\ band.}
\tablenotetext{e}{Dominated by dynamic range.}
\label{tab:vla}
\end{deluxetable}


\begin{deluxetable}{lccccc}
\small
\rotate
\tablecolumns{6}
\tablewidth{0pt}
\tablecaption{VLA 8.7 GHz Continuum Properties$^{\rm a}$}
\tablehead{
\colhead{} & \colhead{$\alpha$} &
\colhead{$\delta$} & \colhead{Peak Flux Density} &
\colhead{Integrated Flux Density} & \colhead{Angular Size FWHM} \\
\colhead{Source} & \colhead{(1950)} &
\colhead{(1950)} & \colhead{(Jy$\, \rm{beam}^{-1}$)} &
\colhead{(Jy)} & \colhead{(arcsec)}
}
\startdata
 
\ngc{6572} & 18~~09~~40.54 & 06~~50~~25.4 & 0.97  & 1.25  & 8.4 $\times$ 9.4 \\
\jon{320}  & 05~~02~~48.35 & 10~~38~~21.0 & 0.018 & 0.024 & 8.3 $\times$ 9.4 \\

\enddata

\noindent
\tablenotetext{a}{Based on two-dimensional Gaussian models.}
\label{tab:cont}
\end{deluxetable}


\begin{deluxetable}{lcccccccc}
\small
\rotate
\tablecolumns{9}
\tablewidth{0pt}
\tablecaption{Planetary Nebula Model Parameters}
\tablehead{

\colhead{} & \colhead{$\theta_{\rm inner}$} & \colhead{$\theta_{\rm outer}$} &
\colhead{$V_{\rm exp}$} & \colhead{$T_{e}$} & \colhead{$n_{e}$} & 
\colhead{} & \colhead{} & \colhead{} \\
\colhead{Source} & \colhead{(arcsec)} & \colhead{(arcsec)} &
\colhead{(\kms)} & \colhead{(K)} & \colhead{(\percc)\expo{3}} & 
\colhead{(\hepr4)} & \colhead{(\heppr4)} & \colhead{(\hepr3)\expo{3}}

}
\startdata
 
\ngc{6572}       & 2.0 &  7.9 & 15.0 & 10300 & 22.0  & 0.11 & 0.00 & 1.0 \\
\jon{320}        & 2.0 &  7.4 & 46.0 &  7500 &  1.5  & 0.10 & 0.00 & 1.9 \\
\jon{320} (halo) & 7.4 & 35.0 & 25.0 &  7500 &  0.05 & 0.10 & 0.00 & 1.9 \\

\enddata

\noindent
\label{tab:model}
\end{deluxetable}

\end{document}